\documentclass[aip, jap, amsmath, amssymb, numerical, preprint, nofootinbib]{revtex4-1} %preprint
\usepackage{graphicx}
\usepackage{dcolumn}
\usepackage{bm}

\begin{document}

\title{Description of the wavevector dispersion of surface plasmon-phonon-polaritons}

\author{V. V. Korotyeyev}
\email{koroteev@ukr.net or vadym.korotieiev@ftmc.lt}
\affiliation{Department of Theoretical Physics, Institute of Semiconductor Physics of NAS of Ukraine,  03028, Kyiv, Ukraine}
\affiliation{Center for Physical Sciences and Technology, Saul\.{e}tekio al. 3, LT-10257, Vilnius, Lithuania}
\author{V. Janonis}
\affiliation{Center for Physical Sciences and Technology, Saul\.{e}tekio al. 3, LT-10257, Vilnius, Lithuania}
\author{I. Ka\v{s}alynas}
\affiliation{Center for Physical Sciences and Technology, Saul\.{e}tekio al. 3, LT-10257, Vilnius, Lithuania}
\begin{abstract}
We reported here the results of the calculations of wavevector dispersion of oscillations frequencies, $\omega'(k)$, and damping $\omega''(k)$ of the surface plasmon phonon polaritons (\mbox{SPPhP})
for the heavy-doped GaN sample. We showed that $\omega'(k)$- dependence consists of the three branches with the specific anticrossing behavior due to the interaction of surface plasmon polariton (SPP) with surface phonon polariton(SPhP).  The strong renormalization of the damping $\omega''(k)$ in the vicinity of the anticrossing region was found.
The obtained dispersions of the $\omega'(k)$ and $\omega''(k)$ were applied for the analytical analysis of exact electrodynamic simulation of the resonant behavior of the reflectivity spectrum of the n-GaN grating.
%the explanations of the specific resonant behavior of the reflectivity spectrum of the GaN sample with surface-relief grating.
\end{abstract}

%\pacs{}% insert suggested PACS numbers in braces on next line
\maketitle

The excitations of surface electromagnetic waves in polar semiconductor material are intensively studied during the past decade with the aims of development of electrically-pumped  thermal emitters with
the strong frequency-selective ({\it coherent}) properties in the terahertz (THz) and far-infrared spectral ranges. The resonant features in reflection, absorption and emission spectra were identified for different structures with the metasurfaces including SiC\cite{Greffet2002}, Ge\cite{Kang2015}, GaAs\cite{Pozela2014,Shaligin2018}, GaN ~\cite{Shaligin2016, Janonis2020}. Recently, the unusual behavior of the surface electromagnetic waves is discussed for 2D-crystalline systems~\cite{Low2017}.The observed features of the THz spectra are attributed to the resonant interactions of light and the surface polaritons with wavevectors determined by the metasurface geometry.

Authors proposed in Ref.~\cite{Janonis2020} new design of the n-GaN grating for the excitation of the \mbox{SPPhP} modes which performed as the resonant features in the reflection and emission spectra.
The SPPhP resonance is realized in the vicinity of the anticrossing point of two hybridized SPP- and SPhP-like branches.
We identified that this resonance, observed in reflectivity spectrum as appearance of the two asymmetric dips, possessing cardinally different dispersion than that of a conventional SPP and SPhP resonances.

In this short communication, we illustrate that strong difference in quality factors of these dips relate to the strong renormalization of the corresponding mode's damping. In the most papers, the study of the SPPhP's dispersion is restricted by analysis of the mode's oscillation frequency neglecting the analysis of their damping.

Here, the dispersion of the SPPhP is calculated for the case of the air/n-GaN interface using the dispersion equation (\ref{Eq1}):
\begin{equation}
k^{2}=\left(\frac{\omega}{c}\right)^2\frac{\epsilon_s\epsilon_0}{\epsilon_s+\epsilon_0},
\label{Eq1}
\end{equation}
where dielectric permittivity, $\epsilon_0$,  corresponds to the air and assumed to be equal 1.  For GaN, we use scalar form of the dielectric function inherent for polar materials (effect of the anisotropy is neglected, here ):
\begin{equation}
\epsilon_s=\epsilon_{\infty} \left( \frac{\omega^2_{LO}-\omega^{2}-i\omega\gamma_{LO}}
{\omega^2_{TO}-\omega^{2}-i\omega\gamma_{TO}}-
\frac{\omega_p^2}{\omega^2+i\omega\gamma_p}\right).
\label{Eq2}
\end{equation}
The first and second contributions describe the polarization of the lattice and electron subsystems, respectively. The $\omega_{LO}$ [$\gamma_{LO}$], $\omega_{TO}$ [$\gamma_{TO}$], $\omega_p$ [$\gamma_p$] are the angular frequencies [damping] of the longitudinal, transversal optical phonons and bulk plasmons, respectively. Below, we assumed that $\gamma_{LO}=\gamma_{TO}\equiv\gamma_{ph}$.  Bulk plasmon frequency, $\omega_{p}=\sqrt{4\pi e^{2}n_{0}/m^{*}\epsilon_{\infty}}$, with $n_{0}$ is the free electron concentration and $\epsilon_{\infty}$ is the high-frequency dielectric permittivity.

\begin{figure}
\includegraphics[width=0.8\textwidth]{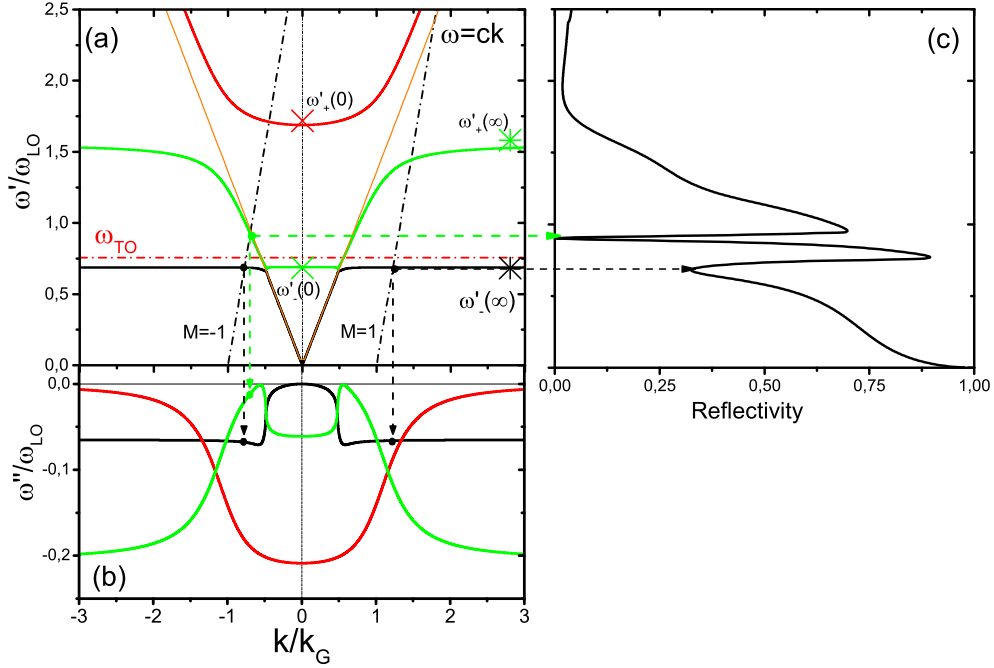}
\caption{The  $\omega'$ (panel a) and  $\omega''$ (panel b) as functions of the dimensionless wavevector. The reflectivity spectrum (panel c) calculated for incidence angle $\theta=26^{o}$ for the surface-relief grating with period of $10$ $\mu m$, grating height of $1$ $\mu m$ and filling factor of $50$ $\%$.  Incline dash-dotted lines are given by phase-matching condition (\ref{Eq3}). Dashed arrows indicate the resonant frequencies and damping of the corresponding modes for which phase-matching is fulfilled. Material parameters of the n-GaN are following: $\omega_{TO}=559$ cm$^{-1}$, $\omega_{LO}=739$ cm$^{-1}$, $\omega_{p}=1130$ cm$^{-1}$, $n_{0}=1.55\times 10^{19}$,  $\epsilon_{\infty}=5.3$; $\gamma_{TO}=\gamma_{LO}=7$ cm$^{-1}$, $\gamma_{p}=390$ cm$^{-1}$.    }
\label{fig1}
\end{figure}

In the calculations of \mbox{SPPhP} dispersion, we assumed the complex-valued frequency, $\omega=\omega'+i\omega''$ ($\omega'$-oscillation frequency and $\omega''$-damping),
while $k$-vector of SPPhP excitation is assumed to be real. This representation can be applied for the characterization of the spectral features of the reflectivity, transmittivity, and absorptivity which are measured at the uniform external illumination (see, for example \cite{Greffet1997} ).

The wavevector dispersions, $\omega'(k)$, and $\omega''(k)$, are shown in the Fig.~\ref{fig1}(a) and (b), respectively.
Fig.~\ref{fig1}(c) provides the reflectivity spectrum of n-GaN grating obtained as a result of exact electrodynamic simulation in the frameworks of Rigorous Coupled-Wave Analysis (RCWA)\cite{Janonis2020}.
Particular parameters of the n-GaN grating are listed in the caption of Fig.\ref{fig1}.

The Eqs.~(\ref{Eq1}) and (\ref{Eq2}) form the algebraic equation of the third order with respect to $\omega^{2}$ and as seen from Fig.~\ref{fig1}(a), there are three branches which are symmetrically located with respect to positive/negative $k$. At small $k$, lowest SPP-branch (black curve) have almost linear dispersion, $\omega'\approx ck$, with negligibly small damping  while highest SPhP-branches (green and red curves) have weak wavevector dispersion with the following asymptotic frequencies:
\begin{align}
\label{Eq4}
\omega'_{\mp}(0)&=\left[\frac{\omega_{LO}^2+\omega_{p}^2\mp\sqrt{(\omega_{LO}^2+\omega_{p}^2)^2-4\omega_{p}^{2}\omega_{TO}^2}}{2}\right]^{1/2}\!.
\end{align}
At large $k$, highest branch (red curve) exhibits the almost linear dispersion, $\omega'\approx ck$,
while black and green branches tend to the following characteristic frequencies:
\begin{align}
\label{Eq5}
\omega'_{\mp}(\infty)&=\left[\frac{\omega_{TO}^2+\epsilon_{\infty}(\omega_{LO}^2+\omega_{p}^2)\mp\sqrt{(\omega_{TO}^2+\epsilon_{\infty}(\omega_{LO}^2+\omega_{p}^2))^2
-4\epsilon_{\infty}(\epsilon_{\infty}+1)\omega_{p}^{2}\omega_{TO}^2}}{2(1+\epsilon_{\infty})}\right]^{1/2}\!.
\end{align}
The frequencies $\omega'_{\mp}(0)$ and $\omega'_{\mp}(\infty)$ for particular parameters of GaN are marked in Fig.\ref{fig1}(a) by crosses and stars, respectively.

As seen, there are regions of $k$ and $\omega$  where SPP and SPhP branches strongly interact. This manifests itself as a emergence of the anticrossing region in $\omega'(k)$-dependencies and strong renormalization of $\omega''(k)$-dependencies (see Fig.~\ref{fig1}(b)). Two modes are strongly coupled in the anticrossing region and there is a sense to introduce the concept of the hybrid modes.

In our case, the hybridization of SPP and SPhP branches occurs at the $\omega'_{-}(0)$ frequency (see Eq.\ref{Eq4}). Note that the coupling of longitudinal-optical (LO) phonons with free-carrier plasmon excitations at  $\omega'_{+}(0)$ frequency has been studied elsewhere (see for example Ref.\cite{Kasic2000}). Far away from the anticrossing points, the branches have a SPP- and SPhP-like behavior with particular damping.

It should be noted that the direct observation of the surface modes in the spectroscopy measurements requires spatially-nonuniform surface. For example, surface-relief grating can provide a efficient coupling between surface waves and incident light. The calculated reflectivity spectrum of the n-GaN grating (see  Fig.~\ref{fig1}(c) ) possess the resonant features which can be attributed to the excitations of the different modes.
The resonant frequencies are given by intersection points of $\omega'(k)$ dependencies with the characteristic lines following from the phase matching condition:
\begin{equation}\label{Eq3}
	k=\frac{\omega'}{c}\sin\theta+Mk_G,
\end{equation}
where $k_{G}=2\pi/P$ (P is grating period), $\theta$-incident angle and $M=\pm 1,\,\pm 2,...$.

As seen, the obtained resonant frequencies (marked by points in Fig.\ref{fig1}(a)) are well agreed with the positions of the resonant dips in the reflectivity spectrum.
These resonances occur in the vicinity of the anticrossing region and they are associated with excitations of the SPhP-like (the broader and lower -frequency resonance) and SPP-like (the narrower and higher-frequency resonance) modes of the $M=-1$ order.

The variations of both grating period (shift of the characteristic lines ) and incidence angles (slope of the characteristic lines) can control the resonant frequencies and facilitate to observe SPPhP resonances. However, it should be noted that dispersion equation (2) is obtained under assumption of the flat surface and it explains the resonances position in the reflectivity spectra only for the considered case of shallow grating. The rigorous treatment of the modes for the relief surfaces requires the proper consideration of the branch folding within the Brillouin zone. Latter is out of the scope of this short communication and will be presented elsewhere.

\end{document}